# Lifespan associations of resting-state brain functional networks with ADHD symptoms

Rong Wang[1,2,3*], Yongchen Fan[3], Ying Wu[3], Yu-Feng Zang[4,5,6], Changsong Zhou[2,7*]

[1]*State Key Laboratory for Strength and Vibration of Mechanical Structures, School of Aerospace Engineering, Xi'an Jiaotong University, Xi'an 710049, China*

[2]*Department of Physics, Centre for Nonlinear Studies, Beijing-Hong Kong-Singapore Joint Centre for Nonlinear and Complex Systems (Hong Kong), Institute of Computational and Theoretical Studies, Hong Kong Baptist University, Hong Kong*

[3] *College of Science, Xi'an University of Science and Technology, Xi'an 710054, China*

[4]*Center for Cognition and Brain Disorders, the Affiliated Hospital of Hangzhou Normal University, Hangzhou, Zhejiang, China*

[5]*Institute of Psychological Sciences, Hangzhou Normal University, Hangzhou, Zhejiang, China*

[6]*Zhejiang Key Laboratory for Research in Assessment of Cognitive Impairments, Hangzhou, Zhejiang, China*

[7]*Department of Physics, Zhejiang University, Hangzhou 310027, China*

*Corresponding authors. Email: wang0712_xust@163.com (R.W.), cszhou@hkbu.edu.hk (C.Z.)

## Abstract

Attention-deficit/hyperactivity disorder (ADHD) is increasingly being diagnosed in both children and adults, but the neural mechanisms that underlie its distinct symptoms and whether children and adults share the same mechanism remain poorly understood. Here, we used a nested-spectral partition (NSP) approach to study the resting-state brain functional networks of ADHD patients ($n$=97) and healthy controls (HCs, $n$=97) across the lifespan (7-50 years). Compared to the linear lifespan associations of brain functional segregation and integration with age in HCs, ADHD patients have a quadratic association in the whole brain and in most functional systems, whereas the limbic system dominantly affected by ADHD has a linear association. Furthermore, the limbic system better predicts hyperactivity, and the salient attention system better predicts inattention. These predictions are shared in children and adults with ADHD. Our findings reveal a lifespan association of brain networks with ADHD symptoms and provide potential shared neural bases of distinct ADHD symptoms in children and adults.

## Introduction

Attention-deficit/hyperactivity disorder (ADHD) is the most common neurological disorder in childhood [1] and is clinically diagnosed with age-inappropriate hyperactivity/impulsivity and inattention. Approximately 40–60% of children with ADHD have persistent symptoms in adulthood, and a recent finding also reported a significant percentage of ADHD in adults [2]. Although adults with ADHD show brain structures and functions different from those of children with ADHD [3-6], their core clinical descriptions are essentially the same [1]. Meanwhile, due to clinical heterogeneity and subjective psychiatric diagnoses [7, 8], it is still challenging to accurately diagnose ADHD [9]. The lifespan exploration of the neural mechanisms of ADHD and linking neural signatures to clinical symptoms are promising approaches for developing more objective and individual-specific diagnoses.

In a worldwide meta-analysis on brain anatomies across the lifespan (4-63 years) [10], ADHD patients were found to have smaller volumes in several regions than healthy controls (HCs), such as the accumbens, amygdala and hippocampus [10]. These anatomical alterations were only apparent in children and disappeared in adults, which suggests a maturation delay during childhood [10-16]. However, Samea et al. found no significant alterations in the regional activation level [17], and whether a delay of maturation in brain functional organization in children parallels anatomical immaturity is still controversial. For example, functional integration (i.e., global cooperation between different systems) in normal brain networks is positively correlated with age [18-20], but both decreased and increased integration have been reported in children with ADHD relative to HCs [21-24]. Meanwhile, it also remains unclear how brain functional organization correlates with age in adults with ADHD. The above questions require a lifespan exploration of functional brains in ADHD patients. In a frontocentral event-related potential (ERP) study, ADHD patients (18-59 years) had a quadratic correlation between NoGo P3 amplitude and age, different from the linear correlation in HCs [25]. It is thus suspected that the brain functional organization of ADHD patients may also have a quadratic association with age across the lifespan.

Hyperactivity and inattention are the major clinical symptoms of ADHD, and these symptoms are thought to have different neural bases [23]. Sudre et al. observed that persistent inattentional symptoms are tied to anomalous connectivity in the default mode network (DMN) [26]. Sanefuji et al. found that the symptoms of the hyperactive subtype of ADHD are related to the corticostriatal network, whereas the symptoms of the inattentive subtype of ADHD are associated with the right ventral attention network [27]. However, as age and ADHD symptoms jointly affect brains [28], the lifespan association of brain functional organization with age is supposed to be affected by ADHD, and the corresponding dominant ADHD effects are thus expected to signify the underlying neural bases for hyperactivity or inattention. Meanwhile, children and adults with ADHD show different brain functions relative to HCs [3, 4], but whether they share the same mechanisms of hyperactivity and inattention is still unknown.

To address the above questions, neural signatures that link the brain to ADHD symptoms across the lifespan need to be extracted. Normal brain functions depend not only on the sufficiently segregated processing in specialized systems but also on the effective global integration among them [29]. Functional segregation and integration in brain functional connectivity (FC) networks have been shown to be reliable biomarkers for cognitive functions [30], and their abnormalities have been linked to brain disorders [29, 31, 32], including ADHD [33]. Thus, it is expected that the segregation/integration features may be associated with ADHD symptoms across the lifespan. However, the graph measures of segregation and integration (e.g., modularity and the participant coefficient) are based on the modular partition at a single level in brain networks [34], which does not allow the detection of segregated and integrated processing across multiple scales. Recently, we developed a nested-spectral partition (NSP) method to detect hierarchical modules in brain networks according to the eigenmodes and described segregation and integration across multiple levels [35]. Hierarchical segregation and integration have been demonstrated to be better neural signatures of cognitive functions than classical signatures [36, 37]. We thus expected that an NSP-based analysis could better reveal the neural biomarkers that underlie distinct ADHD symptoms across the lifespan.

Therefore, in this work, we studied hierarchical segregation and integration in brain FC networks and explored lifespan associations with distinct ADHD symptoms. Hierarchical modules in FC networks were analyzed using resting-state functional magnetic resonance imaging (fMRI) datasets of children and adults with ADHD and HCs with a wide range of ages (7-50 years). We first extracted the lifespan association of brain FC networks with age in the ADHD and HC groups and studied the alterations of network segregation and integration related to ADHD in different age ranges. Second, we identified the dominant effects of age and ADHD on different functional systems and investigated their heterogeneous functional patterns across the lifespan. Finally, we tested whether brain systems differentially affected by ADHD or age could selectively predict distinct ADHD symptoms and whether these predictions are specific in ADHD patients relative to HCs.

## Results

The data for 97 ADHD patients and 97 age/sex-matched HCs were extracted from three centers, and the clinical scores for hyperactivity, inattention and total symptoms were collected to describe the severity of ADHD symptoms [38]. Resting-state FC networks ($N$=100 regions) were constructed for each participant using the Pearson correlation coefficient [39] and were further multisite corrected (see Materials and Methods). Functional segregation and integration components (i.e., $H_{Se}$ and $H_{In}$) were computed using the NSP method [35]. At the whole-brain level, $H_{Se}$ and $H_{In}$ were negatively correlated across the subjects in both groups (Fig. S1), and a higher $H_{Se}$ or smaller $H_{In}$ reflected stronger network segregation. Since the shorter length of an fMRI series biased the network to more segregation [37, 40], group-averaged segregation and integration components were calibrated to the corresponding values of the stable FC network that was constructed by concatenating all fMRI time series of all participants in each group [37]. This combination of concatenation across a long enough time and calibration generated the fMRI length-independent network measures for all participants in each group and has been found to be advantageous in linking the brain to cognitive abilities [37]. Pertinently, calibrated segregation and integration components for each region (i.e., $H_{Se}^{i}$ and $H_{In}^{i}$, $i=1\cdots N$) were also extracted to reflect the regional contribution to overall network segregation and integration (see Materials and Methods for details).

## Quadratic lifespan association of the brain functional network with age in ADHD patients

The likelihood ratio test (LRT) was used to identify the lifespan associations between brain network segregation/integration and age (Table S1), and the bootstrapping statistics also provided similar results (Table S2). In the HCs, we found a linear association between brain functional organization and age (Fig. 1a). Across the lifespan (7-50 years), the global integration component $H_{In}$ was positively correlated with age ($p$=0.048), and the segregation component $H_{Se}$ was negatively related to age ($p$=0.019, see Fig. 1a), which indicates increased network integration on the global scale of the normal brain with age, and this is consistent with the previous result that used single-level module detection [18]. However, the ADHD patients had a typically quadratic lifespan association with age in brain FC networks (Fig. 1b). The integration component first increased with age and then decreased after approximately 30 years of age. This quadratic relationship is significant (age$^2$: $p$=0.045, see Fig. 1b). Meanwhile, the segregation component first decreased with age and then increased, which is also significant (age$^2$: $p$=0.023). Furthermore, we divided each group into three age-binned subgroups roughly termed childhood (CH, 7-19 years), adulthood (AH, 20-35 years) and old adults (OA, 36-50 years). In the HCs, the OA subgroup had the highest connectivity density in FC networks, and CH and AH had nearly the same density (Figs. 1c and S2), consistent with the positive correlation of network integration with age. In ADHD patients, the AH subgroup had the highest connectivity density compared with the CH and OA subgroups that further manifested the first rising and then declining patterns of network integration with age. Therefore, on the global scale, the resting-state brain functional network in ADHD patients had an abnormally quadratic association between network integration and age, which is different from the linear relationship in HCs.

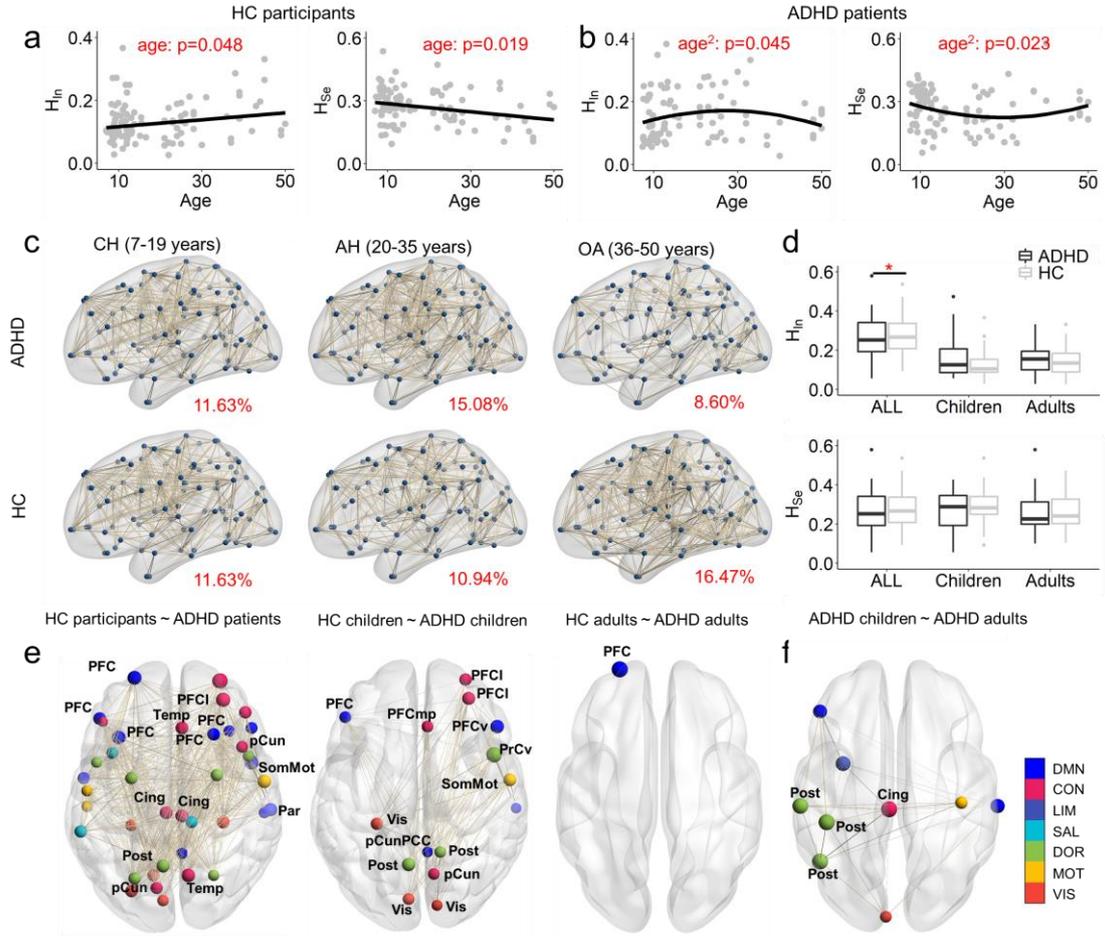

**Fig. 1. Abnormal lifespan associations between brain functional networks and age in ADHD patients.** Lifespan associations of network segregation and integration components with age in (a) HC participants and (b) ADHD patients. These fitting models were determined by LRT and bootstrapping (see Tables S1 and S2). (c) Averaged FC networks for different subgroups with different age ranges visualized using BrainNet Viewer [41] with a binarizing threshold of 0.55. The connectivity densities were provided (see Fig. S2 for more comparisons with other binarizing thresholds). (d) Comparisons of the network integration component $H_{In}$ and segregation component $H_{Se}$ between the ADHD and HC groups in all participants (ALL), children (7-19 years) and adults (20-50 years). * MANOVA $p<0.05$. (e) (f) Visualizations of the subnetworks in different comparisons. These regions had significant alterations in the integration component or segregation component ($p<0.05$), and they formed subnetworks. A larger node size represents a higher increase in the degree (total FC to the node) of the weighted subnetwork, and a thicker edge indicates a higher increase in FC. Regions were colored according to their belonging to different systems, and those marked with regional names had significantly increased degrees within the subnetworks (*p*<0.05). In the adults with ADHD, only one region was detected in the DMN system, which was also robust in a separate analysis of adult data from one site (see Fig. S3). *DMN - default mode network; LIM - limbic; SAL - salient attention; DOR - dorsal attention; VIS - visual; CON - control; MOT - somatomotor. PFC - prefrontal cortex, Cing - cingulate, Post -posterior, pCun - precuneus, SomMot - somatomotor, PFCl - lateral prefrontal cortex, Temp - temporal, Par – parietal, Vis - visual, pCunPCC - precuneus posterior cingulate cortex, PFCv - ventral prefrontal cortex, PrCv - precentral ventral.*

## ADHD-related network alterations in children and adults

Previous works have reported inconsistent effects of ADHD on brain network segregation and integration in children or adults [18, 21-24]. When taking all participants into consideration, ADHD patients had a higher integration component on the global scale (Fig. 1d, *p*=0.026) but an insignificant alteration in the segregation component (*p*=0.432). Notably, the inverted U-like association of functional organization with age in ADHD patients implies different alterations of brains in children and adults. In children, ADHD patients had a higher integration component and smaller segregation component (*p*=0.078 and 0.442, Fig. 1d), and these alterations on the global scale were significant in a separate analysis of data from the two sites (multisite corrected, Fig. S4). Similarly, adults with ADHD had a higher integration component and smaller segregation component than HCs, and these alterations were insignificant (*p*=0.172 and 0.766, see Fig. S3b). There was also no significant difference between ADHD children and ADHD adults in the integration component (*p*=0.717) and segregation component (*p*=0.265).

Thus, the alterations in ADHD patients may be located in local regions. In all participants, the regions with significant alterations of $H_{In}^{i}$ and $H_{Se}^{i}$ related to ADHD were mainly located in the control and DMN systems (all *p*<0.05, uncorrected, Fig. 1e). More importantly, most of these regions did not show a significant ADHD-related alteration if we considered connectivity degrees in the whole-brain FC network (Fig. S5). However, while a subnetwork was formed by these regions with significantly altered integration or segregation components, we found that the regions with a significantly increased degree of connectivity *within the subnetwork* related to ADHD were distributed in the control and DMN systems (Fig. 1e, *p*<0.05). With the same procedure, we defined the subnetworks for children wherein the regions had significantly altered integration components or segregation components (Fig. 1e, *p*<0.05). The significant regions have an increased degree, and they are distributed in the control, dorsal attention, DMN and visual systems. Only one significant PFC region was detected in the comparison between ADHD adults and HC adults, which was in the DMN system, and it maintained robust changes in a separate one-site analysis (Fig. S3). Furthermore, we also identified the subnetwork in the comparison between ADHD adults and ADHD children (Figs. 1f and S6). These significant regions in ADHD adults had a higher contribution to functional integration than in ADHD children (*p*<0.05, uncorrected) and a higher degree in the subnetwork. Nearly all significantly different regions between ADHD adults and ADHD children were located in the dorsal attention and control systems.

Overall, ADHD-related hyperconnectivity across the lifespan was mainly found in local regions located in the DMN and control systems, but children and adults had more specific alterations. The abnormalities in children were mainly located in the control, dorsal attention, DMN and visual systems, but they were located in the DMN in adults. Crucially, children with ADHD and adults with ADHD had significant differences in their dorsal attention and control systems.

**Heterogeneous effects of ADHD and age on brain functional organization**

We next investigated the dominant effects of ADHD and age on functional systems. Using a multiple-regression approach (see Materials and Methods), we evaluated the effect of age and the effect of ADHD on the segregation/integration components ($H_{In}$ or $H_{Se}$) in each functional system. In all patients, age and ADHD had heterogeneous effects on different functional systems (Fig. 2a). For the network integration component, age had the largest negative effect on salient attention and motor systems, and ADHD had the largest effect on the limbic system (Fig. 2a). In terms of the network segregation component, age had the largest positive effect on the salient attention system, and ADHD had the largest effect on the limbic system (Fig. 2a). While performing a principal component analysis (PCA) on the effects of age and ADHD on network integration and segregation components, we obtained an overall coeffect defined as the difference between the first component for $H_{In}$ (explaining 86.4% of the variance) and the first component for $H_{Se}$ (explaining 80.6% of the variance). A larger positive coeffect indicates a higher effect of

ADHD on brain network integration, and a larger negative coeffect represents a higher effect of age. It is clear to see a higher effect of ADHD on the limbic system and a higher effect of age on the salient attention and motor systems (Fig. 2b). However, if we performed the analysis separately for the children and adult subgroups, then this coeffect exhibited a great difference between children with ADHD and adults with ADHD. In ADHD children, age had the largest effect on the dorsal attention system, but the effect of age was in the salient attention system for ADHD adults. Meanwhile, ADHD had a high coeffect on the limbic system in both children and adults.

We found that the heterogeneous effects of ADHD and age on functional systems in children and adults relate to different lifespan functional patterns. All systems had similar quadratic lifespan associations with age in the integration and segregation components (Figs. 2c and S7), except for the limbic system, which was statistically tested by LRT and bootstrapping (Table S1). The quadratic lifespan associations of FC with age were mainly located around the salient attention and control systems (Table S3). Meanwhile, the fitting line of the $H_{In}$ of the limbic system in ADHD patients was above that for HCs (Fig. 2c), but this difference in the fitting lines between the ADHD and HC groups was insignificant ($p=0.243$). We thus further compared the segregation/integration of this system at CH (7-19 years), AH (20-35 years) and OA (36-50 years) between the two groups and found that AH ADHD patients had significantly higher integration than AH HCs ($p<0.05$, see Fig. S8). Thus, even though the limbic system has a similar linear lifespan association with age in ADHD patients and HCs, ADHD-related increased integration indeed exists.

Therefore, although age and ADHD jointly affect the brain's resting state in patients, the limbic and salient attention systems relate to different effects across the lifespan. Children and adults with ADHD share a dominant effect of ADHD on the limbic system that has a linear lifespan association with age; however, age dominantly affects the salient attention system in adults but affects the dorsal attention system in children.

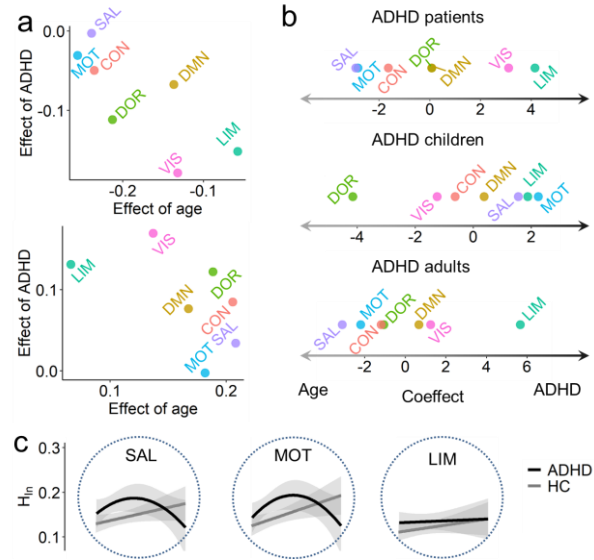

**Fig. 2. Heterogeneous lifespan associations between functional systems and age in ADHD patients.** **(a)** Effect of age and effect of ADHD on network integration (upper panel) and segregation components (lower panel) in different functional systems. **(b)** PCA-based overall coeffect between age and ADHD on brain network integration for all ADHD patients, ADHD children and ADHD adults. **(c)** Lifespan associations of $H_{In}$ with age in three typical systems with large coeffects (see Fig. S7 for $H_{Se}$). These curves were obtained by fitting the $H_{In}$ of HC and ADHD participants with age, and the fitting models were determined by LRT and bootstrapping. The shadow indicates the confidence interval. The linear fitting models in the limbic system were first obtained, and the average difference of the predicted values from the fitted models between ADHD and HC groups was calculated. Then, the permutation test (1000 times) was applied to obtain a distribution of the average differences in a null model, and the *p*-value was provided.

**The limbic system better predicts hyperactive symptoms in ADHD patients**

While functional systems were heterogeneously affected by ADHD and age and had different lifespan associations with age in the ADHD patients, we expected that these heterogeneous lifespan functional patterns signify distinct mechanisms of hyperactivity or inattention. To test this possibility, linking resting-state brain network properties to ADHD symptoms is urgently needed. In addition to the network integration and segregation components $H_{In}$ and $H_{Se}$, we further measured the heterogeneity of regional segregation/integration components (i.e., $CV_{In}$ and $CV_{Se}$) since the brain requires the heterogeneous activation of certain regions to achieve task switching

[3]. The heterogeneities were calculated for the whole brain and all functional systems. The highly negative correlation between $CV_{In}$ (or $CV_{Se}$) and $H_{In}$ (or $H_{Se}$) indicates that brain networks with higher integration/segregation correspond to a more homogeneous distribution of the regional integration/segregation component (Fig. 3a).

We performed multiple linear regression models while controlling for sex and age, and a beta estimation was used to represent the correlation between the ADHD scores and brain measures (see Materials and Methods). In all ADHD patients, the $H_{In}$ of the limbic system had the highest correlation with the hyperactive score (see Fig. 3b, $\beta$=-0.276, $p$=0.020). The negative correlation implies higher hyperactivity for less network integration. Meanwhile, the $CV_{In}$ of the visual system was positively correlated with the hyperactive score ($\beta$=0.254, $p$=0.033), which indicates higher hyperactivity for a more heterogeneous distribution of the regional integration component, matching to less network integration. Thus, it seems consistent that the limbic and visual systems dominantly affected by ADHD can better predict hyperactivity in ADHD patients.

However, there is another possibility that the limbic system can better predict the hyperactive score in both ADHD children/adults and HCs. In the children and adults with ADHD, we also found that a higher hyperactive score was related to less network integration (Figs. 3c, 3d). The limbic system had the highest correlation between $CV_{In}$ and the hyperactive score in children with ADHD ($\beta$=0.368, $p$=0.044) and between $H_{In}$ and the score in adults with ADHD ($\beta$=-0.416, $p$=0.002). In addition, we further collected the hyperactive score of healthy children ($n$=26, 8-16 years, data not available for healthy adults) and used different brain measures to predict it. Contrary to the ADHD children, healthy children had a positive correlation between the hyperactive score and network integration, and the best prediction was not in the limbic system (Fig. 3e). Therefore, the limbic system better predicts hyperactivity in ADHD patients, which is closely related to its dominant effect of ADHD but is independent of age.

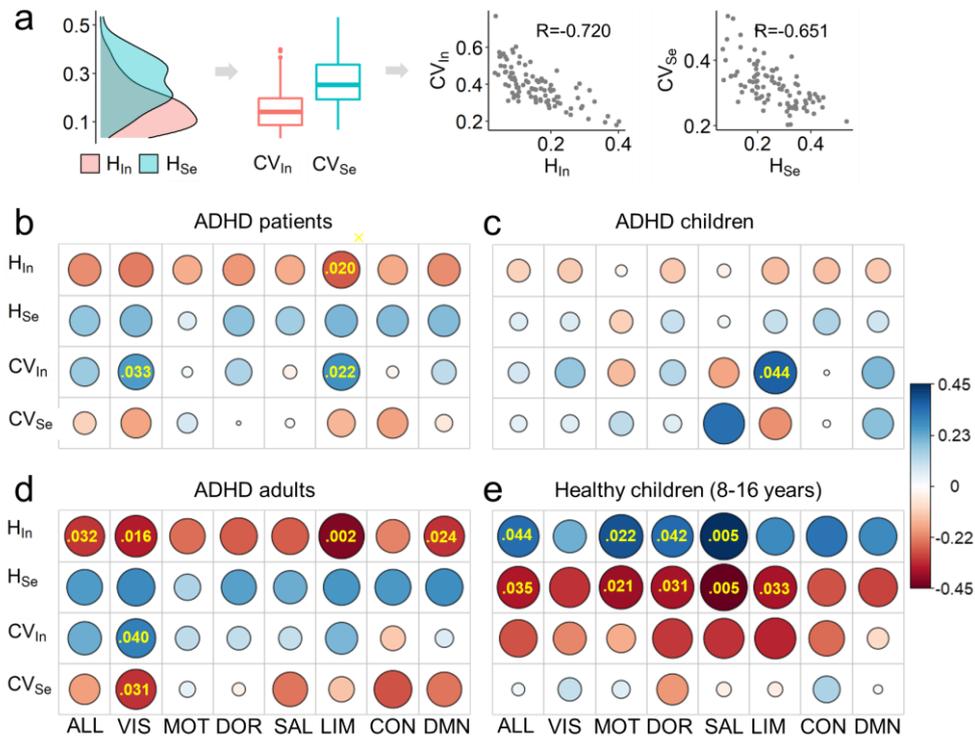

**Fig. 3. The limbic system better predicts hyperactivity in ADHD patients**. (a) Definitions of $CV_{In}$ and $CV_{Se}$ measuring the spreading of the regional $H_{In}$ and $H_{Se}$ (left panel) and their correspondences to the integration and segregation components in the whole brain. (b) Beta estimations measuring the correlations between the hyperactive scores and brain measures in ADHD patients, (c) ADHD children, (d) ADHD adults and (e) healthy children for the whole-brain (ALL) networks and seven functional systems. The significant predictions were provided along with the *p*-values.

**The salient attention system better predicts inattention in ADHD patients**

Similar to hyperactivity, we next tested whether there is a special system that can better predict inattentive scores and whether this system is specific in ADHD patients. In ADHD patients, the $CV_{Se}$ of the salient attention system was significantly related to the inattentive score (*β*=0.233, *p*=0.047, Fig. 4a), which indicates that higher inattention is associated with a more heterogeneous distribution of the regional segregation component. In children with ADHD, the $CV_{Se}$ of the salient attention system also had the highest correlation with the inattentive score (*β*=0.555, *p*=0.016, see Fig. 4b). Importantly, the salient attention system also better predicted the inattentive scores in adults with ADHD (*β*=-0.394, *p*=0.007, see Fig. 4c), and the negative correlation between $CV_{In}$ and the inattentive scores indicates higher inattention for a more homogeneous distribution of the regional integration component. However, in healthy children, the inattentive score had a positive correlation with the $CV_{In}$ of systems (Fig. 4d), contrary to that in ADHD children. The salient attention system cannot predict the score. Since the salient attention system does not have a consistent dominant coeffect in ADHD adults and children, these results indicate that the salient attention system that better predicts inattentive severity in ADHD patients was a specific property relative to HCs and was independent of the coeffects of age and ADHD.

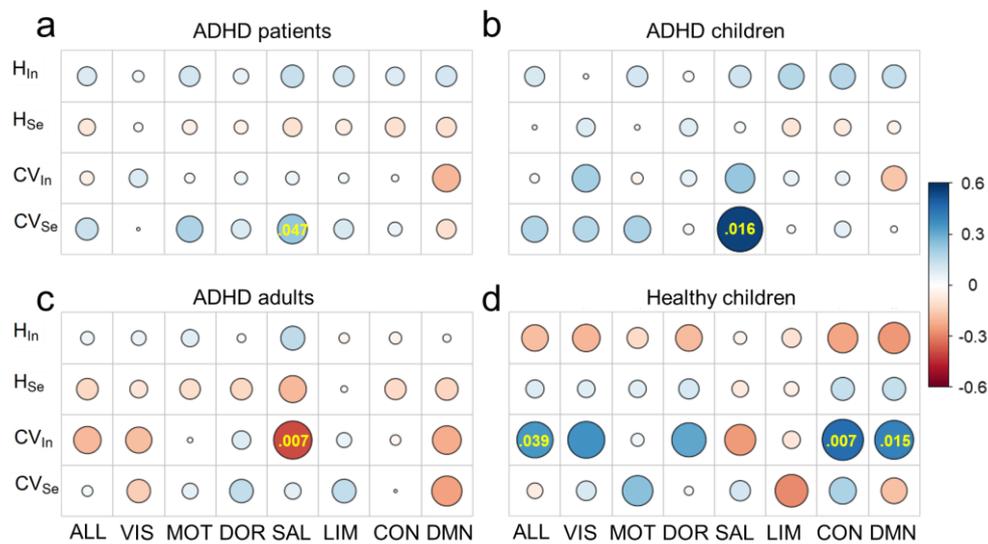

**Fig. 4. The salient attention system better predicts inattentive scores in ADHD patients.** Beta estimations between the hyperactive score and brain measures in **(a)** ADHD patients, **(b)** ADHD children, **(c)** ADHD adults and **(d)** healthy children for the while-brain (ALL) networks and seven functional systems. The significant predictions were provided along with the *p*-values.

## Discussion

To link the brain functional organizations with ADHD clinical symptoms across the lifespan, we measured functional segregation and integration based on hierarchical modules in brain FC networks. We first found a quadratic lifespan association of brain FC networks with age in ADHD patients. Second, we showed that ADHD was related to abnormal hyperconnectivity of local regions in the DMN and control systems across the lifespan, and the abnormal regions were mainly located in the control system for children and in the DMN for adults. Compared to ADHD children, ADHD adults had higher integration in several regions that were mainly located in the dorsal attention and control systems. Third, the limbic system was dominantly affected by ADHD in both children and adults, and this system had a linear lifespan association with age. However, age dominantly affected the dorsal attention system in children with ADHD and the salient attention system in adults with ADHD. Finally, we found that the limbic system better predicted hyperactivity, and the salient attention system better predicted inattention. These predictions were consistent and shared between ADHD children and adults. Our results reveal the abnormal lifespan associations of brain functional networks with age in ADHD patients and provide the potential distinct neural bases of hyperactive and inattentive symptoms.

Age has complex effects on the segregation and integration of resting brain functional organizations, such as increased network integration with enhanced average FC (4-7 years) or decreased FC (6-10 years) with age [19, 42]. Several studies have reported that elderly individuals exhibit higher integration than younger individuals [18, 20], but decreased integration was also reported [43]. Another study found that network segregation increases during childhood development and peaks in young adulthood [44]. Here, we found a significantly positive linear correlation between age (7-50 years) and network integration in HCs, providing further evidence for the increase of brain network integration with age [18, 20], which may be accompanied by an increase in crystallized intelligence and a decrease in fluid intelligence [45]. In children with ADHD (7-16 years), previous studies have found a decrease in local FCs within the DMN with age [46, 47], but the FCs in HCs showed inconsistent relations with age [46, 47]. Meanwhile, when using an independent component analysis (ICA), a component loading appeared to decrease with age in children with ADHD (8-15 years), while it appeared to become greater in HCs [5]. In adults with ADHD (21-60 years), the FC within the executive control network decreased with age [48]. These cross-sectional and local FC explorations are not sufficient to identify the manner in which both age and ADHD affect the network segregation and integration of resting-state brains on a global scale. Here, we found that brain FC networks have a quadratic correlation with age in ADHD patients across the lifespan relative to the linear association in HCs. Thus, our work offers the first lifespan evidence that network integration first increases and then decreases with age in ADHD patients. Furthermore, this result may be consistent with the ERP result, where ADHD patients (18-59 years) had a quadratic association of NoGo P3 amplitude with age, different from the linear relationship in HCs [25]. On the other hand, a worldwide lifespan meta-analysis reported the delayed maturation of brain volumes in children with ADHD but insignificant structural alterations in adults with ADHD [10]. Our results further indicate that the functional alterations may not parallel the structural abnormalities in ADHD patients.

Across the lifespan, ADHD has different effects on brain FC networks in children and adults. Generally, ADHD has been hypothesized to be a DMN-dysconnectivity disorder [26, 49-53], which embraces the abnormalities of the DMN in ADHD and its return to normal functioning after treatment with methylphenidate [54]. Indeed, aberrant FC within the DMN was present in children and adults with ADHD [4, 51], but the alterations were inconsistent in children with ADHD [23, 46, 55, 56] and adults [3, 56-58]. An insignificant connectivity change within the DMN was also observed in children with ADHD [50]. These inconsistencies in ADHD children and adults may be related to the medication, course of disease, severity, hyperactive/inattentive subtypes, etc. [59]. However, in the lifespan study, a meta-analysis combining children and adults with ADHD reported significantly altered FCs distributed in the DMN and control systems [51], and we also found that ADHD patients had functionally abnormal regions within the DMN and control systems, and these regions have increased integration contribution (or degree) compared to HCs. Our results partially match those of previous studies [23, 46, 60], but notably, here, the subnetworks were extracted based on the regional alterations of segregation and integration at the whole-brain scale, which is different from the predefined subnetwork analysis. Since the DMN is highly active during rest but becomes deactivated during task performance [61, 62], the DMN hypothesis proposed that due to poor deactivation during tasks [55], the DMN is less able to effectively transition from a baseline to an active state [63]. Our results imply that the hyperconnected DMN at rest in ADHD patients lost its segregation ability to flexibly transition to task states. Meanwhile, the control system plays a key role in regulating the functions of other networks [64] and is associated with ADHD-related mind wandering [65] and symptom remission [66]. In a longitudinal follow-up study, persistent ADHD was related to higher FC within the control system, which was further increased for remitting ADHD [66]. Here, we found that in all ADHD patients, the control system had regions with significantly increased integration contribution (or degree), but this was not related to ADHD symptoms. This higher integration may compensate for the ADHD deficit [67] and may be an efficient mechanism to suppress ADHD symptoms [66].

Even though a previous study reported that children with ADHD and adults with ADHD shared altered FCs within the DMN and between the DMN and ventral attention network [4, 8, 28, 68], we did not find any shared abnormal regions. According to neurodevelopmental theory [69], ADHD remission is driven by improved prefrontal top-down control. A longitudinal follow-up study reported that increased FC within the control system corresponds to less severe ADHD symptoms [66]. Here, the control system could not predict clinical symptoms in ADHD patients. The abnormal regions were located in the DMN system in ADHD adults and were distributed in the control, dorsal attention and visual systems in ADHD children. Our results suggest the neural mechanism transition of ADHD from widespread abnormalities in children to more concentrated abnormalities in adults. These results also indicate the intrinsic difference between ADHD adults and ADHD children. A meta-analysis showed that hypoactivation in the motor system during tasks was less prominent in ADHD adults than in ADHD children [3], which is in line with the decrease in hyperactivity with age [70]. Compared to ADHD children, we found a smaller inattentive score in ADHD adults ($p=0.013$) and a higher integration contribution of regions in the dorsal attention and control systems during rest. Thus, enhanced executive control functions may contribute to the remission of ADHD symptoms.

Children/adolescents (7.2–21.8 years) with ADHD were found to have a functional maturation lag in the DMN [28], and young ADHD rats (4-6 weeks) had a lag in limbic regions [71]. Here, we found that the limbic system was significantly different between the two groups during AH (20-35 years), but across the lifespan, this system was predominantly affected by ADHD and showed a linear lifespan association with age that was insignificantly different from that in HCs. This apparent insignificant change in the lifespan association may be partially related to the small sample size and partition of the limbic system that has the smallest number of voxels, and most of them lie in areas likely to be contaminated with susceptibility artifacts. From the perspective of cognitive function, the limbic system involves a set of regions in the paleocortex, which supports a variety of functions related to emotion regulation and motivation meditation and has been known to be associated with ADHD [4, 10, 71, 72]. The normal development of limbic circuitry underlies the reduction in impulsive choices from early adolescence to mid-adulthood [73], and the immature limbic system confidently predicts hyperactivity [12, 14, 74, 75]. Sanefuji et al. also found that the hyperactive subtype is related to the corticostriatal network that is involved to some extent in limbic cortices [27]. However, whether the functional pattern of the limbic system is closely correlated with hyperactivity across the lifespan is unclear. Here, we demonstrate that the limbic system dominantly affected by ADHD can better predict hyperactivity in ADHD patients. This result provides further knowledge that abnormalities in the limbic system also underlie the increase in hyperactive choice across the lifespan [74].

The salient attention system (also called the ventral attention system) was dominantly affected by age in ADHD adults but not in ADHD children. However, this system can better predict inattention in both children and adults with ADHD and was not related to the inattentive scores in healthy children. This result indicates that the salient attention system is closely related to the inattentive score uniquely in ADHD patients rather than in HCs. Meanwhile, the predictions revealed that brains with more homogeneous integration components or more heterogeneous segregation component distributions in the salient attention system correspond to higher inattention. Indeed, to achieve task switching, the brain needs to activate certain regions of the salient attention system and suppress others [3], which may generate higher heterogeneity in the integration component. Thus, our results indicate that a more homogeneous integration component or a more heterogeneous segregation component in the salient attention system at rest may contribute to inefficient task switching that requires the manipulation of attention. From the perspective of cognitive function, the salient attention system was thought to enable brains to direct attention toward salient stimuli by excluding irrelevant noise, which supports automatic "bottom-up" forms of attention [3, 76]. The dysfunction of the salient attention system was thus believed to cause attention deficits related to ADHD [3, 4, 7, 8, 15, 51, 63]. For example, compared to the combined and hyperactive subtypes of ADHD, the predominantly inattentive subtype is more specifically related to an abnormal salient attention system [77], such as increased FC in the right salient attention system [27]. Meanwhile, the salient attention system is a typical task-positive network that modulates the dynamic switching between the DMN and control systems [78]. Abnormal communications among the salient attention, DMN and control systems may induce inattention [23, 63]. Thus, even though we did not observe significant changes in the salient attention system related to ADHD, the significantly abnormal DMN and control systems may contribute to the close mapping between the salient attention system and inattention in ADHD patients. In particular, children with ADHD had abnormalities in the control system, but adults with ADHD had abnormal DMN. Our results may suggest discriminative neural mechanisms of inattention in children and adults with ADHD, wherein inattentive symptoms are indirectly driven by abnormalities in the control system in children but indirectly driven by the DMN in adults.

Accordingly, a hierarchical module analysis enabled the discovery of functional systems that revealed heterogeneous lifespan associations with age and robustly predicted the hyperactive and inattentive symptoms of ADHD patients. The identified functional circuits provide insight into the neurobiological mechanisms that support the important clinical components of ADHD shared in children and adults, which may, in turn, have implications for the development of more objective and accurate diagnostic standards and contribute to the ability to distinguish between the hyperactive and inattentive ADHD subtypes.

## Materials and methods

**Dataset.** The data for 57 children with ADHD and 57 healthy children were extracted from the Peking University Center and New York University (NYU) Child Study Center in the ADHD-200 project (Table 1). The data for 40 ADHD adults and 40 healthy adults were collected from the University of California, Los Angeles (UCLA) project [38]. The data selection principle is provided in the Supplementary Materials. In the Peking and UCLA datasets, the ADHD Rating Scale IV (ADHD-RS) was used to evaluate the clinical scores of hyperactivity/impulsivity, inattention and total symptoms, and in the NYU data, the Conners' Parent Rating Scale-Revised, Long version (CPRS-LV) was used to obtain the ADHD scores. Here, the ADHD-RS scores were used to study the relationship between brain networks and ADHD symptoms. Adults with ADHD had smaller total symptom scores and inattentive scores than children with ADHD (MANOVA, $p=0.044$ and $0.013$), and there was an insignificant difference in hyperactivity ($p=0.614$).

Table 1. Demographic, clinical and neuropsychological features of ADHD patients and healthy

controls.

|  | ADHD-200 (age 7-19) | | UCLA (age 20-50) | |
|  | ADHD | HC | ADHD | HC |
| --- | --- | --- | --- | --- |
| N/female | 57/18 | 57/30 | 40/20 | 40/20 |
| Age | 10.78±2.37 | 10.72±2.25 | 32.05±10.41 | 31.28±9.23 |
| Hyperactivity | 21.81±6.33 | 15.40±3.84 | 21.12±4.58 | |
| Inattention | 26.75±5.19 | 18.54±3.85 | 24.32±2.76 | |
| Total | 48.17±6.18 | 33.94±6.04 | 45.45±4.81 | |
| FD | 0.14±0.05 | 0.14±0.11 | 0.15±0.09 | 0.16±0.12 |

**MRI data processing.** The MRI data of these participants had the same repetition time [TR] = 2 s, and more detailed scanning parameters are provided in the Supplementary Materials. An analysis of Functional NeuroImages (AFNI) (http://afni.nimh.nih.gov/afni/) and the FMRIB Software Library (FSL) (http://www.fmrib.ox.ac.uk/fsl/) were used to preprocess the resting-state fMRI data [35, 79]. The mean framewise displacement (FD) was significantly smaller than the suggested value (0.3 mm) [80], and there was no significant difference in the FD between the ADHD and HC groups (two-sample t-test, $p$=0.605). Echoplanar imaging (EPI) images were motion- and slice-time corrected and spatially smoothed using a Gaussian kernel of 6 mm full-width at half-maximum (FWHM). The fMRI signal was further filtered with a bandpass of 0.01 Hz $< f <$ 0.1 Hz. Additionally, several sources of nuisance covariates were eliminated using a linear regression as follows: 1) 6 rigid body motion correction parameters and 2) the signal from the white matter and the signal from a ventricular region of interest. The global whole-brain signal was not removed.

**Resting-state brain FC.** We used the Schaefer atlas, which is based on the transitions of FC patterns [39], to parcellate the brain into $N$=100 regions of interest (ROIs). This resolution of the atlas has also been used in a recent ADHD study [24]. The BOLD signals of voxels belonging to one region were averaged to obtain the regional fMRI data. To overcome the effect of different lengths on the results, the length of the BOLD signal was controlled to be the same and lasted for 304 s (152 frames). The Pearson correlation coefficient was used to compute the FC between any two regions. Here, stable FCs within groups and individual static FCs were separately constructed. First, the fMRI time series were concatenated among all participants in each group, and stable FCs were obtained. Second, for each participant, the total fMRI series was used to construct the individual static FC. Finally, the negative correlations in the FC matrices were set to zero, and the diagonal elements were kept at one. Here, the mean percentage of positive connectivity in the individual FC matrices in the HC group was 91.86% and was 93.33% in the ADHD group. Following previous studies [24, 37], negative connectivity was excluded. This operation also contributes to clarifying the statistical relationship between brain networks and ADHD symptoms (see Fig. S9 for more discussion).

**Harmonization of multisite datasets.** Our datasets were extracted from three different centers; thus, the multisite effect should be properly considered. ComBAT software was used to harmonize the static FC [81]. In this software, there are mainly two control setting parameters, namely, the batch vector and biological variables. The batch vector specifies the scanner of the data, and biological variables indicate the information that should be protected during the removal of scanner effects, i.e., sex, ADHD diagnosis and age in this study.

**Hierarchical modules of FC networks.** The NSP method was applied to identify the segregation and integration of brain FC networks based on eigenmodes [36, 37]. Using eigendecomposition, eigenvectors $U$ and eigenvalues $\Lambda$ of FC matrix C were sorted in descending order of $\Lambda$. NSP first detected the hierarchical modules of the FC networks with the following procedures (see Fig. S10):

1. The 1$^{st}$ mode had the same sign of eigenvector values for all regions and was regarded as the first level with one module (i.e., whole-brain network).

2. In the 2$^{nd}$ mode, the regions with positive eigenvector signs were assigned to a module, and the remaining regions with negative signs formed the second module. This mode was regarded as the second level with two modules.

3. According to the positive or negative eigenvector sign of the regions in the 3$^{rd}$ mode, each module in the second level could be further partitioned into two submodules to form the third level. Successively, the FC network could be partitioned into modules of multiple levels as the order of functional modes increased. When each module contained only a single region at a given level, the partitioning process was stopped. Additionally, the regions within a module at a specific level may have the same sign of eigenvector values in the next level; then, the module was indivisible, which had no effect on the subsequent iterative process. During the partitioning process, the module number $M_i(i=1,\cdots,N)$ and modular size $m_j(j=1,\cdots,M_i)$ at each level were recorded.

**Hierarchical segregation and integration in brain FC networks.** Different from classical segregation and integration based on modules at a single level [34], the hierarchical segregation and integration components of brain FC networks were defined across multiple levels [37]. The first level in the FC network had a single large module, which corresponded to the global network integration with the largest eigenvalue $\Lambda$. The second level with two modules supported the local integration within each module and the segregation between them, which required a decreased eigenvalue. With an increasing mode order, more modules reflected deeper levels of the segregated process, accompanied by smaller eigenvalues $\Lambda$. The segregation and integration components at each level can be defined as [37]

$$H_i = \Lambda_i^2 M_i (1 - p_i)/N \quad (1)$$

with

$$p_i = \frac{\sum_j |m_j - N/M_i|}{N}. \quad (2)$$

Here, $N$ is the number of regions, and $p_i$ is a correction factor for the heterogeneous modular size and reflects the deviation from the optimized modular size $m_j = N/M_i$ in the $i$-th level. The global integration component is thus taken from the first level:

$$H_{In} = H_1/N, \quad (3)$$

and the segregation component is summed from the $2^{nd}$ - $N^{th}$ levels:

$$H_{Se} = \sum_{i=2}^{N} H_i/N. \quad (4)$$

At the whole-brain level, $H_{Se}$ and $H_{In}$ were negatively correlated across subjects in both groups (Fig. S1). Based on the orthogonal and standard eigenvectors, the network integration and segregation components in each level could be mapped to each region $j$:

$$H_{In}^j = H_1 U_{1j}^2 \text{ and } H_{Se}^j = \sum_{i=2}^{N} H_i U_{ij}^2, \quad (5)$$

where $U_{ij}$ is the eigenvector value for the $j$-th region at the $i$-th level. The segregation and integration of a functional system can be obtained by averaging the corresponding components of the regions in this system. Then, the distributions of the regional segregation/integration components were measured with the coefficient of variance:

$$CV_{In} = \frac{\sigma_{H_{In}^j}}{\overline{H_{In}}} \quad \text{and} \quad CV_{Se} = \frac{\sigma_{H_{Se}^j}}{\overline{H_{Se}}}. \tag{6}$$

Here, $\sigma_{H_{In}^j}$ and $\sigma_{H_{Se}^j}$ are the standard variances among regions across the whole brain or any functional system, and $\overline{H_{In}}$ and $\overline{H_{Se}}$ represent the corresponding averages. These measures based on NSP are more powerful in linking brain networks to distinct ADHD symptoms than a classical FC analysis (Fig. S11).

**fMRI length calibration.** A proportional calibration strategy was used to overcome the bias of brain FC networks to higher segregation in shorter fMRI series [37, 40]. The group-stable segregation and integration components, i.e., $H_{In}^S$ and $H_{Se}^S$, could be calculated from each stable FC matrix built from concatenated fMRI time series. The vectors of segregation (or integration) components from individual static FC networks for all participants in each group are $H_{In} = [H_{In}(1), H_{In}(2), \cdots, H_{In}(97)]$ and $H_{Se} = [H_{Se}(1), H_{Se}(2), \cdots, H_{Se}(97)]$, which were calibrated to $H_{In}^{'}(n) = H_{In}(n) \times H_{In}^S / \langle H_{In} \rangle$ and $H_{Se}^{'}(n) = H_{Se}(n) \times H_{Se}^S / \langle H_{Se} \rangle$ for the *n*-th participant. Here, $\langle \rangle$ represents the average across all participants. This calibration was separately performed in each group. Then, the calibration of regional segregation and integration was also performed. For region *j* of the *n*-th participant, the calibrated segregation and integration components are $H_{Se}^{j'} = H_{Se}^j / H_{Se}(n) \times H_{Se}^{'}(n)$ and $H_{In}^{j'} = H_{In}^j / H_{In}(n) \times H_{In}^{'}(n)$, where the relative contribution of each region to network segregation/integration remained consistent.

**Effects of age and ADHD.** We built different multiple-regression models to obtain the effects of age and ADHD on the brain [28]. In all patients, the regression model was

$$H = \beta_1 \times age^2 + \beta_2 \times ADHD + \beta_3 \times age + \beta_4 \times sex + \beta_5 \times FD + \varepsilon. \tag{7}$$

Here, $H$ is the brain measure, and $\varepsilon$ is the residual. In this model, the brain measures were affected by age, ADHD symptoms, sex and head motion (FD). The parameter $\beta_1$ measures the effect of age, and $\beta_2$ stands for the effect of ADHD. To maintain consistency, this model was also applied to the limbic system even though it had a linear lifespan association with age (see

Tables S1 and S2).

In children or adults with ADHD, the network segregation and integration components were linearly related to age. Thus, the regression model was

$$H = \beta_1 \times age + \beta_2 \times ADHD + \beta_3 \times sex + \beta_4 \times FD + \varepsilon \qquad (8)$$

This model does not consider the nonlinear effect of age on brain functional organization. The above models were separately fitted for $H_{In}$ and $H_{Se}$ in each functional system. Thus, each model has the $\beta_1$ and $\beta_2$ series for each measure in seven systems. Then, a PCA of these estimation coefficients was performed, and the subtraction difference between the first components for integration and the segregation components was obtained to measure the coeffect of age and ADHD on the participants' brains.

## Statistical tests

The linear regression model $y \sim x + FD$ and quadratic regression model $y \sim x^2 + x + FD$ were applied to fit the lifespan association with age. The LRT was used to identify which model was chosen. If the *p*-value of LRT was smaller than 0.05, then we chose the quadratic regression model; otherwise, the linear regression model was used. Bootstrapping also provided similar results. A multivariate analysis of covariance (MANOVA) was used to assess the alterations induced by ADHD in Figs. 1c-f, controlling for sex, age and FD. A linear regression model was conducted to examine the relationships between distinct ADHD symptoms and brain measures in Figs. 3 and 4. These statistical tests were performed in *R*.

## Code and data availability

The original datasets are available at https://openneuro.org/datasets/ds000030 and http://fcon_1000.projects.nitrc.org/indi/adhd200/index.html. The brain atlas and the partition of seven functional systems are available at https://github.com/ThomasYeoLab/CBIG/tree/master/stable_projects/brain_parcellation. The ComBAT is available at https://github.com/Jfortin1/ComBatHarmonization. The codes used in this study are available at https://github.com/TobousRong/ADHD.

# Acknowledgments

This work was supported by the National Natural Science Foundation of China (Nos. 11802229, 11972275, and 11772242), the Hong Kong Scholars Program (No. XJ2020007), and the Hong Kong Baptist University (HKBU) Research Committee Interdisciplinary Research Cluster Matching Scheme 2018/19 (IRCMS/18-19/SCI01). This research was conducted using the resources of the High Performance Computing Cluster Centre, HKBU, which receives funding from the RGC, University Grant Committee of Hong Kong and HKBU.

# Competing interests

The authors declare no competing interests.